\documentclass[amsmath,amssymb,10pt,aps,notitlepage,prl,twocolumn,superscriptaddress,nofootinbib,longbibliography]{revtex4-1}
\usepackage[colorlinks=true]{hyperref}
\usepackage{lipsum}
\usepackage{soul}
\usepackage{graphicx}
\usepackage{latexsym}
\usepackage{amsmath}
\usepackage{amsthm}
\usepackage{amssymb}
\usepackage{epstopdf} 
\usepackage{enumitem}
\usepackage{setspace}
\usepackage{dcolumn}
\usepackage{bm}
\usepackage{setspace} 
\usepackage{slashed}
\usepackage{color}
\usepackage{youngtab}
\usepackage{tikz}
\usepackage{tikz-3dplot}
\usepackage{braket}
\usepackage{cancel}
\usepackage{subfigure}
\usepackage{pbox}
\usepackage[normalem]{ulem}

\usepackage{lipsum}

\usetikzlibrary{shapes,snakes,arrows,chains,matrix,positioning,scopes,calc}
\usetikzlibrary{decorations.markings}
\usepackage{moreverb}
\allowdisplaybreaks 
\usepackage{blindtext}

\begin{document}

\title{Preparation of Entangled Many-Body States with Machine Learning}
\author{Donggyu Kim}
\author{Eun-Gook Moon}
\email{egmoon@kaist.ac.kr}
\affiliation{Department of Physics, Korea Advanced Institute of Science and Technology (KAIST), Daejeon 34141, Korea}
\date{\today}

\begin{abstract}
	Preparation of a target quantum many-body state on quantum simulators is one of the significant steps in quantum science and technology. 
	With a small number of qubits, a few quantum states, such as the Greenberger-Horne-Zeilinger state, have been prepared, but fundamental difficulties in systems with many qubits remain, including the Lieb-Robinson bounds for the number of quantum operations. 
	Here, we provide one algorithm with an implementation of a deep learning process and achieve to prepare the target ground states with many qubits.
	Our strategy is to train a machine-learning model and predict parameters with many qubits by utilizing a pattern of quantum states from the corresponding quantum states with small numbers of qubits. 	For example, we demonstrate that our algorithm with the Quantum Approximate Optimization Ansatz can effectively generate the ground state for a 1D XY model with 64 spins (energy error $\simeq0.154\%$). We also demonstrate that the reduced density operator of two qubits can be utilized to capture the pattern of quantum many-body states such as correlation lengths even for quantum critical states.
\end{abstract}
\maketitle

\begin{figure}[t]
	\centering
	\includegraphics[width=0.71\columnwidth]{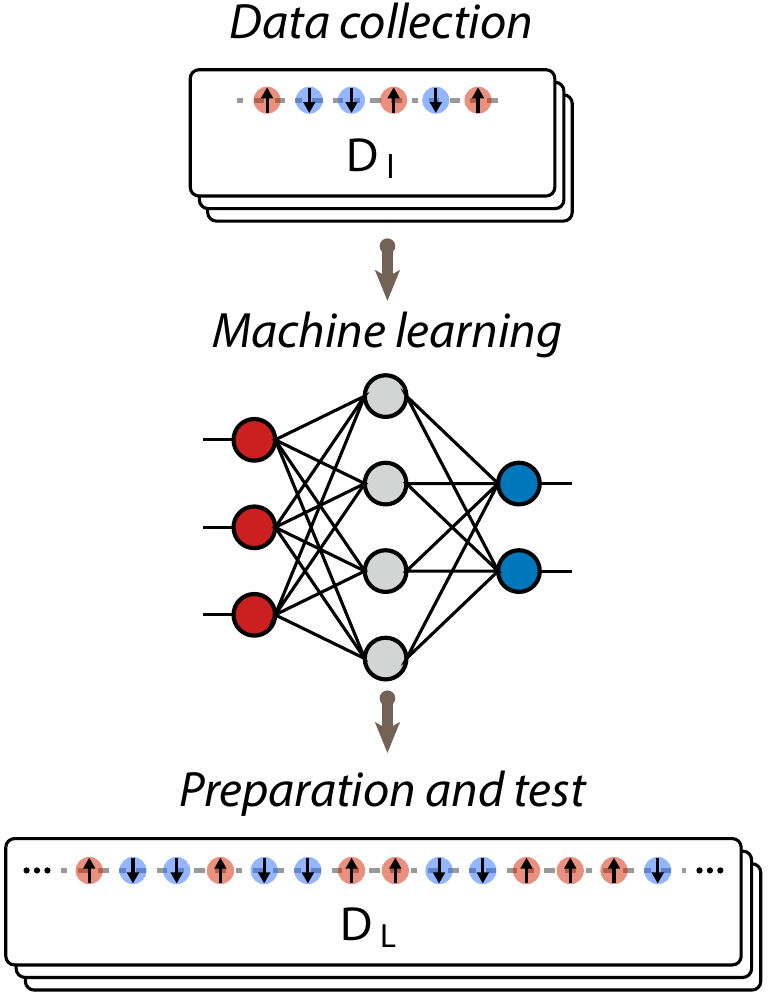} 
	\caption{Schematic overview of a machine-learning implemented hybrid quantum-classical algorithm for preparing ground states for large systems. By running a hybrid quantum-classical algorithm, data $D_l$ are gathered for the ground state with a small number of qubits ($l$).  A machine learning model is then trained using the data for small numbers of qubits $\{D_{l_1},D_{l_2},\cdots\}$, allowing it to identify patterns in ground states. Using the machine learning model, the data $D_L$ for large number of qubits ($L$) are predicted, and these are used to generate the ground state with large number of qubits.}
	\label{Fig:1}
\end{figure}

\textit{Introduction.} Significant advancements in quantum many-body systems have been achieved in synthetic quantum systems, such as trapped ions \cite{blatt2012quantum,zhang2017observation,islam2013emergence}, ultracold atoms \cite{greiner2002quantum,bloch2008many,bernien2017probing}, and superconducting qubits \cite{devoret2013superconducting,gambetta2017building}. 
For example, the preparation of the Greenberger-Horne-Zeilinger (GHZ) state has been achieved \cite{monz201114}, and non-trivial entanglement structures of quantum many-body systems such as dynamics of black holes have been reported \cite{sachdev1999quantum, calabrese2004entanglement,dvali2014black,sachdev2009quantum, jafferis2022traversable}.
Several variational proposals to simulate a target quantum state, so-called hybrid quantum-classical algorithms, have been proposed, including Hardware efficient ansatz, unitary coupled cluster ansatz, and Quantum Approximate Optimization Ansatz (QAOA) \cite{kandala2017hardware, romero2018strategies, farhi2014quantum,zhou2020quantum, ho2019efficient, kim2021advancing}. 
Each proposal has advantages and disadvantages, but their good performances have been demonstrated for systems with a small number of qubits.

Yet, for systems with a large number of qubits, direct applications of the proposals are severely limited due to the exponential growth of a Hilbert space with the number of qubits. 
For example, one has to deal with a $2^{40}\times2^{40}$ matrix to optimize variational energy with 40 qubits in QAOA, which is numerically challenging.
Furthermore, it is well known that highly entangled states require the number of local unitary operations proportional to a system size \cite{ho2019efficient}. 
Thus, an algorithm applicable to systems with many qubits is strongly called for to prepare a target state with many qubits. 

\begin{figure*}[t]
	\centering
	\includegraphics[width=2.0\columnwidth]{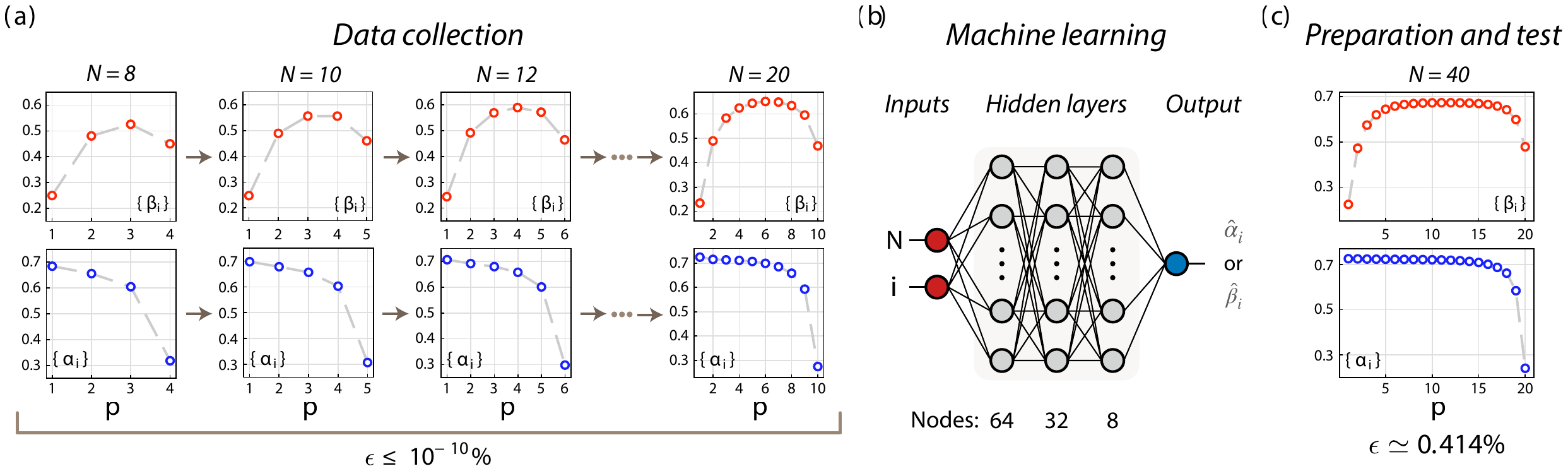} 
	\caption{Efficient parameter prediction for TFIM. (a) Optimal parameters $\{\alpha_i, \beta_i\}$ for variational ansatz Eq.\,\eqref{eq_Ising_ansatz} to generate ground states of varying system sizes. We employ the classical Broyden-Fletcher-Goldfarb-Shanno optimization to obtain optimal parameters $\{\alpha_i,\beta_i\}$ for small systems (system size of 8 to 20). The energy errors $\epsilon$ for the prepared states with the optimal parameters are less than $10^{-10}\%$. (b) A graphical representation of neural networks used for $\{\alpha_i\}$ and $\{\beta_i\}$. We use the system size $N$ and parameter number $i$ as inputs. The neural network consists of 3 hidden layers, which are specified by the weights and biases. The output of the neural network for $\{\alpha_i\}$($\{\beta_i\}$) corresponds to a predicted parameter $\hat{\alpha}_i$($\hat{\beta}_i$) for system size of $N$. (c) Optimal parameters $\{\alpha_i,\beta_i\}$ for a system size of 40 are predicted by the neural networks trained on the parameters for small systems, which have learned the patterns in the parameters of the ground states. The prepared state with the predicted parameters exhibits an energy error $\epsilon$ in order of $0.414\%$.}
	\label{Fig:Ising process}
\end{figure*}

In this work, we provide such an algorithm and demonstrate our success in preparing a target quantum state with more than 60 qubits. Like numerous other studies that have employed machine learning in the field of physics \cite{carleo2017solving, gao2017efficient, torlai2018neural, glasser2018neural, choo2018symmetries, carrasquilla2017machine, ch2017machine, broecker2017machine, zhang2017quantum, zhang2017machine, hsu2018machine, zhang2018machine, sun2018deep, greplova2020unsupervised, ohtsuki2016deep, balabanov2020unsupervised}, our strategy is to implement machine learning to the previous proposals to prepare ground states for large systems in quantum simulators (see Fig.\,\ref{Fig:1}).
Machine learning is compelling in uncovering a pattern of parameters for any ground states recognized based on the ground states for small systems and predicting parameters for large systems.
We demonstrate that the ground states for the transverse-field Ising model with 40 qubits are well prepared, and their validity is checked by small energy error from the exact values and entanglement entropy for quantum critical points. 
Our results indicate that the target ground states for large systems can be constructed using our algorithm in near-term quantum simulators.

\textit{Machine learning with hybrid quantum-classical algorithms.} 
\\\noindent We propose a machine-learning implemented hybrid quantum-classical algorithm to prepare the ground state of a Hamiltonian with the following three steps. 
\begin{enumerate}
	\item Data collection: \\ to run a hybrid quantum-classical algorithm and collect data ($D_{l}$) on the ground state with a small number of qubits ($l$).  
	\item Machine learning:\\ to train a machine learning model with $\{D_{l_1}, D_{l_2}, \cdots\}$.
	\item Preparation and test: \\ to generate the data ($D_{L}$) for a system with a large number of qubits based on the machine learning model and check its validity.
\end{enumerate}
We stress that the three steps of our algorithm provide a general guideline, and their details must be applied case by case.

The machine learning model and hybrid quantum-classical algorithm are the two main ingredients of our algorithm. 
In this work, as a proof of principle, we choose a simple neural network for the machine learning model and QAOA for the hybrid quantum-classical algorithm, even though we believe that our algorithm is generically used with different machine learning models and hybrid quantum-classical algorithms. 

A few remarks are as follows. First, the ground state preparation with many qubits (for example, 64 qubits) is highly challenging with any hybrid quantum-classical algorithms. Not only the number of quantum operators but also the classical optimization cannot be done generically. Our algorithm overcomes the difficulty by utilizing machine learning.  
Second, the efficiency of our algorithm depends on the choice of the machine learning model and the hybrid quantum-classical algorithm. Our results, such as the error of the prepared ground state's energy, become its upper bound. 
Below, we demonstrate our algorithm with three prime examples. 

\textit{Transverse Field Ising model.} 
Let us consider the transverse field Ising model (TFIM) in a chain whose Hamiltonian is defined as 
\begin{eqnarray}
	H_I(g) = -(1-g)\sum_{i=1}^N \sigma^z_i\sigma^z_{i+1}-g\sum_{i=1}^N\sigma^x_i, 
	\label{Ising model}
\end{eqnarray}
where $\sigma^z_i(\sigma^x_i)$ are for the Pauli operators at a site $i$. The periodic boundary condition is imposed, and a dimensionless coupling constant $g$ is introduced. 
The model is well known to have a quantum phase transition at $g = 1/2$ \cite{sachdev1999quantum}, where the ground state is characterized by its conformal symmetry properties. 
The emergent conformal symmetry at $g=1/2$ indicates a non-trivial structure of quantum entanglement of the ground state. The ground state preparation with quantum circuits requires the number of local unitary operations proportional to $N$ with QAOA, indicating the ground state is highly entangled. 

Below is how our algorithm is applied to the TFIM.
\begin{enumerate}
	\item  \{TFIM\} Data collection with QAOA  with the initial state $|\{+\}\rangle$ $\rightarrow$  \{$D_{8}, D_{10}, \cdots, D_{20}$\}. 
	\item \{TFIM\} Machine learning with a 2-64-32-8-1 architecture and a sigmoid activation function  
	\item  \{TFIM\} The ground state preparation ($N=40$) 
\end{enumerate}

The quantum operation part of QAOA is done by the unitary operator alternation with $H_1 = \sum_i\sigma^x_i$ and $H_2 = \sum_i \sigma^z_i\sigma^z_{i+1}$ with variational parameters ($\{\alpha_i,\beta_i\}$), giving the variational many-body state,%
\begin{eqnarray}
	|\psi\left(\{\alpha_i,\beta_i\}\right)\rangle_p = \prod_{i=1}^p\left(e^{i\alpha_iH_1}e^{i\beta_iH_2}\right)|\{+\}\rangle.
	\label{eq_Ising_ansatz}
\end{eqnarray}
The unentangled initial state $|\{+\}\rangle$ is the ground state of $H_1$, and the subscript $p$ is for the depth of QAOA. 
The classical part of QAOA is done by minimizing the variational energy, 
\begin{eqnarray}
	E_p\left(\{\alpha_i,\beta_i\}\right) = {}_p\langle\psi\left(\{\alpha_i,\beta_i\}\right)|H_I|\psi\left(\{\alpha_i,\beta_i\}\right)\rangle_p, \nonumber
\end{eqnarray}
which guarantees the preparation of the ground state. In literature, its efficiency and precision have been demonstrated, showing that the ground states of any $g$ are perfectly prepared when the depth becomes half of the system size, $p = N/2$ up to the system with 18 qubits \cite{ho2019efficient}.   
We stress that not only the quantum part but also the classical part become much more expensive whenever the number of qubits increases. 

To bypass the difficulties with many qubits, we implement machine learning to QAOA. 
First, we collect the data $\{\alpha_i,\beta_i\}$ for  the systems with $N=8, 10, \cdots, 20$, by using the Broyden-Fletcher-Goldfarb-Shanno (BFGS) optimization. We find that the energy errors from the exact values are less than $10^{-10}\%$, which are illustrated in Fig.\,\ref{Fig:Ising process} (a) (see Supplementary Material (SM) for details).
Second, the neural networks are trained with the data for systems with $N=8, 10, \cdots, 20$ with the mean absolute error as the loss function. We used the Adam optimizer with a learning rate of $0.001$, $\beta_1 = 0.9$, and $\beta_2 = 0.999$. We calculated energy errors for the systems with $N = (22, \cdots, 28)$ to evaluate our trained model's performance. 
Third, we generate the parameters for $N = 40$ with the well-trained neural networks as illustrated in Fig.\,\ref{Fig:Ising process} (c) and evaluate the energy errors. We find that the energy error is similar to $0.414\%$.

Furthermore, we also evaluate the entanglement entropy for each subsystem with a size $n$ and then fit it to the exact result, 
\begin{eqnarray}
	S(n) = \frac{c}{3}\ln{\left(\frac{N}{\pi}\sin{\frac{\pi n}{N}}\right)} + A, \label{entanglement_entropy}
\end{eqnarray}
where $c$ indicates the central charge and $A$ is a non-universal constant \cite{calabrese2004entanglement}. 
Our estimation gives  $c = 0.497$, only 0.6$\%$ error from the exact conformal field theory (CFT) prediction value $c = 0.5$. 
Therefore, we demonstrate our algorithm successfully prepares the ground state of TFIM with a large number of qubits (for example, $N=40$) with a small error. 

\begin{figure}[t]
	\includegraphics[width=0.85\columnwidth]{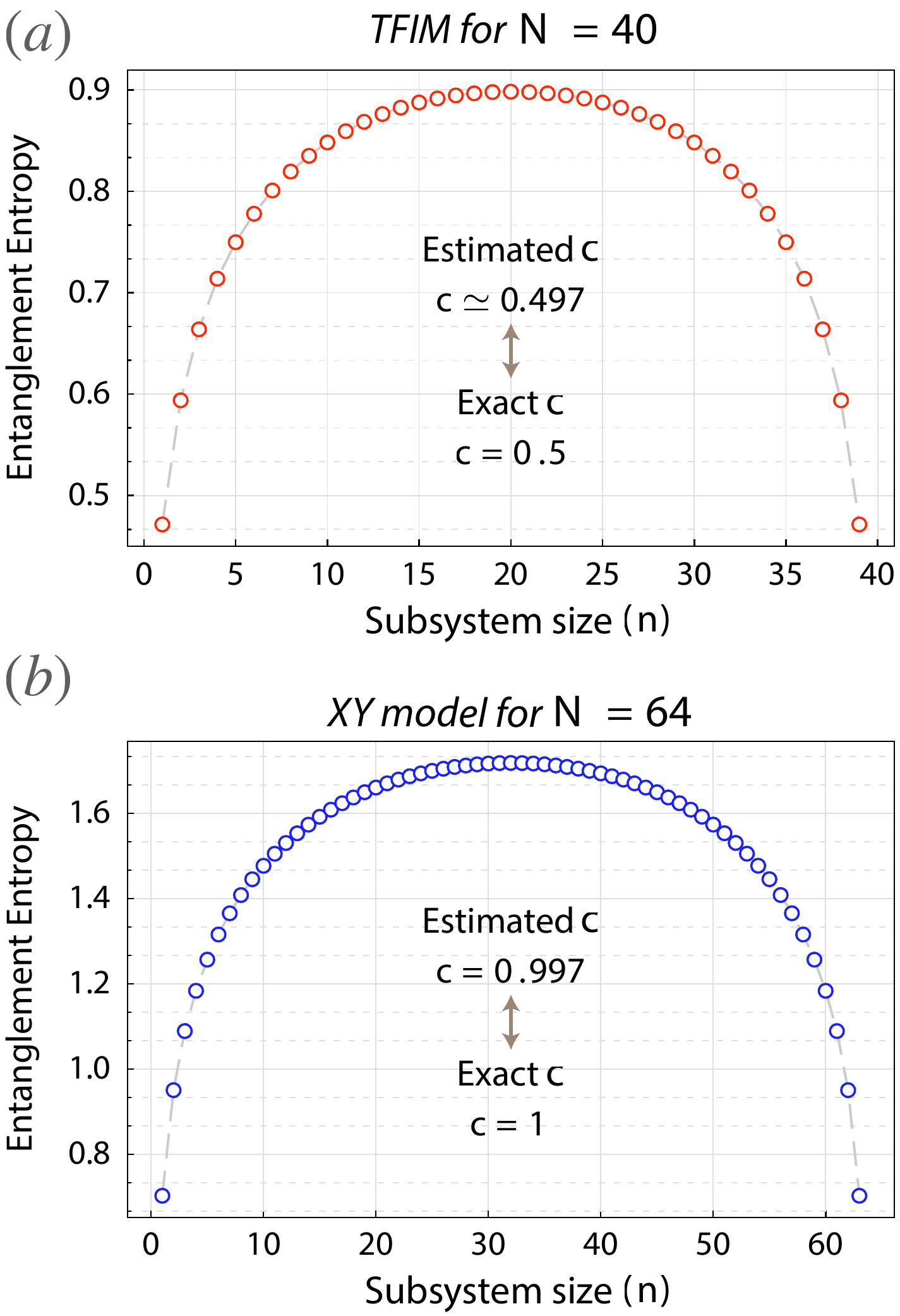} 
	\caption{
		Entanglement entropy values of states prepared by the machine learning-based algorithm as a function of subsystem size $n$. (a) Entanglement entropy values for 40 spins TFIM. The entanglement entropy values match well with the function Eq.\,\eqref{entanglement_entropy}, and the central charge $c$ estimated by fitting is 0.497, which is close to the CFT prediction of 0.5. (b) Entanglement entropy values for 64 spins XY model. The entanglement entropy values are also well fitted to the function Eq.\,\eqref{entanglement_entropy}, and the estimated central charge $c$ is 0.997, comparable to the CFT prediction of 1.
	}
	\label{Fig:Ising and XX}
\end{figure}

\textit{XY model and XXZ model.} We next consider the ground state of the XY Hamiltonian,
\begin{eqnarray}
	H_{XY} = -\sum_i^N \big( \sigma_i^x\sigma_{i+1}^x + \sigma_i^z\sigma_{i+1}^z \big),
	\label{XYeq}
\end{eqnarray}
whose ground state is gapless with the entanglement entropy Eq.\,\eqref{entanglement_entropy}. 
We impose the periodic boundary condition with $4k\, (k\in\mathbb{N})$ qubits. 

Our algorithm is applied to the XY model as follows.
\begin{enumerate}
	\item  \{XY\} Data collection with QAOA  with the initial state $|GHZ\rangle$ $\rightarrow$  \{$D_{16}, D_{20}, \cdots, D_{40}$\}. 
	\item \{XY\} Machine learning with a 2-64-32-8-1 architecture and a sigmoid activation function. 
	\item  \{XY\} The ground state preparation ($N=64$).
\end{enumerate}
The quantum operation part of QAOA is done by the unitary operator alternation with $H_1 = \sum_{i=1}^N \sigma^z_i\sigma^z_{i+1}$ and $H_2 = \sum_{i=1}^N \sigma^x_i\sigma^x_{i+1}$ with variational parameters ($\{\alpha_i,\beta_i\}$), giving the variational many-body state,%
\begin{eqnarray}
	|\psi\left(\{\alpha_i,\beta_i\}\right)\rangle_{N/4} = \prod_{i=1}^{N/4}\left(e^{i\alpha_iH_1}e^{i\beta_iH_2}\right)|GHZ\rangle.
	\label{eq_XX_ansatz}
\end{eqnarray}
Choosing the initial state as the Greenberger-Horne-Zeilinger state ($|GHZ\rangle$) is beneficial, 
\begin{eqnarray}
	|GHZ\rangle = \frac{1}{\sqrt{2}}\left(\otimes_i |\uparrow \rangle_i + \otimes_i |\downarrow \rangle_i \right). \nonumber
\end{eqnarray}

To implement our algorithm, we find that the exact mapping between the TFIM and the XY model is useful, which allows us to obtain optimal parameters for ground states without explicit calculations. 
Roughly speaking, the XY model can be mapped to the doubled TFIM (see SM for details), which makes the central charge twice as it should be. 
Then, by training the machine learning model with the data for $N=16, 20, \cdots 40$, we prepare the ground state of the XY model with 64 qubits. Its energy error is about $0.154\%$, and the fitting to the entanglement entropy relation Eq.\,\eqref{entanglement_entropy} gives the estimation $c=0.997$, which is only $0.3\%$ error to the exact value $1$.

We also analyze the ground state of the XXZ model with even qubits and open boundary conditions, 
\begin{eqnarray}
	H_{XXZ}(g) = -\sum_{i=1}^{N-1}\big( \sigma_i^x\sigma_{i+1}^x + \sigma_i^z\sigma_{i+1}^z+g \sigma_i^y\sigma_{i+1}^y \big).
	\label{eq_XXZ}
\end{eqnarray}
This model becomes the XY model for $g=0$ and the Heisenberg model for $g=1$, where the latter can be exactly solvable by using the Bethe ansatz  \cite{lu2009quasiparticles}. 
Specifically, we choose $g=0.1$ even though our algorithm may apply to a generic $g$.

The application steps of our algorithm are similar to the one of TFIM, and we only comment on the differences below.
First, we find that the quantum operation part of QAOA may be effectively done by the unitary operator alternation with  $H_{o}$ and $H_{e}$ with variational parameters ($\{\alpha_i,\beta_i\}$), giving the variational many-body state,
\begin{eqnarray}
	|\psi\left(\{\alpha_i,\beta_i\}\right)\rangle_{N/2+1} = \prod_{i=1}^{N/2+1}\left(e^{i\alpha_iH_{o}}e^{i\beta_iH_{e}}\right)|\varphi_0\rangle, 
	\label{eq_XXZ_ansatz}
\end{eqnarray}
where $H_{o}$($H_{e}$) refers to the sum of the odd(even) site terms in Hamiltonian Eq.\,\eqref{eq_XXZ}.
We find that the choice of the initial state 
\begin{eqnarray}
	|\varphi_0\rangle = \otimes_i\frac{1}{\sqrt{2}}\left(|\uparrow\uparrow\rangle + |\downarrow\downarrow\rangle\right)_{2i-1,2i}, \nonumber
\end{eqnarray}
is particularly useful. 
Second, the model is not exactly solvable for $g \neq 0$. Instead, one can employ the DMRG  method to find the ground state and its energy, which can be easily done with the open boundary condition. 
For $N=10,12,\cdots 24$, our QAOA procedure gives the overlaps between the prepared states and the ground states from DMRG are larger than $0.998$ (see SM for details).
By employing the neural networks, we uncover the predicted parameters $\{\alpha_i,\beta_i\}$ for $N=40$ and find that the prepared ground state has the overlap ($0.990$) with the DMRG ground state.

\begin{figure}[t]
	\centering
	\includegraphics[width=0.9\columnwidth]{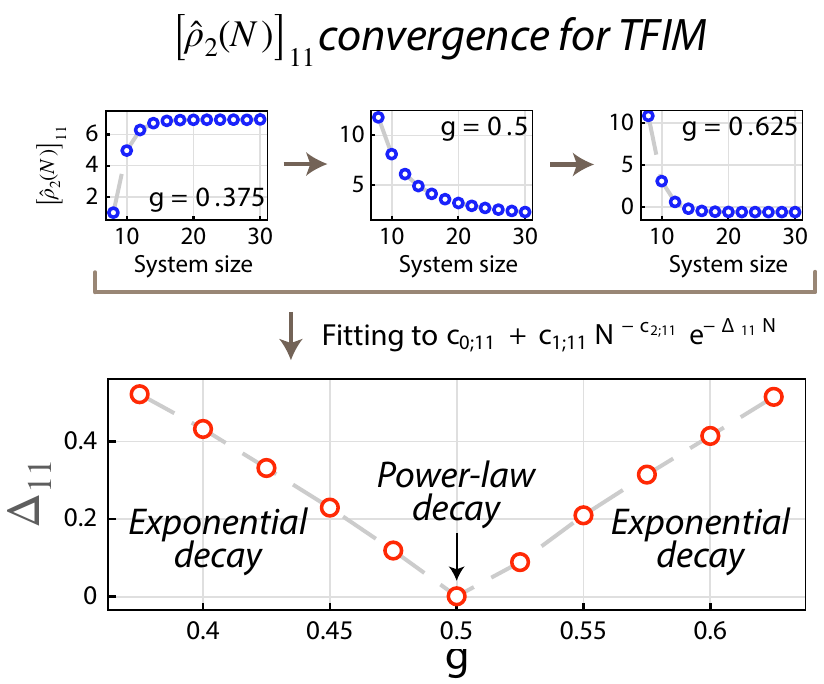} 
	\caption{Analysis of convergence behaviors for reduced density operators of ground states for TFIM. The reduced density operators of two qubits, denoted as $\hat{\rho}_2 (N)$, are computed for different qubit numbers $N$, and the $(1, 1)$th elements $[\hat{\rho}_2(N)]_{11}$ (blue plots) are fitted to $ c_{0; 11} + c_{1; 11} N^{-c_{2; 11}} e^{-\Delta_{11} N}$. In the blue plots, arbitrary units are used on the vertical axes to express the approximate trend of $[\hat{\rho}_2(N)]_{11}$. $e^{-\Delta_{11}N}$($N^{-c_{2; 11}}$)  signifies the exponential(power-law) decay, and when $\Delta_{11}$ approaches 0, the power-law decay becomes significant.}
\label{reduced matrix convergence}
\end{figure}
%

\textit{Discussion and Conclusion.} 
To capture the pattern quantitatively, we consider the reduced density operator of two qubits for the predicted state, $| \Psi(\{\alpha,\beta\})  \rangle$
\begin{eqnarray}
	\hat{\rho}_2(L) = {\rm Tr}_{N-2}\big( | \Psi(\{\alpha,\beta\})  \rangle \langle \Psi(\{\alpha,\beta\})  | \big), 
\end{eqnarray}
where ${\rm Tr}_{N-2}$  is for the tracing over $N-2$ qubits. We find that a matrix component of $\hat{\rho}_2(N)$ for ground states generically has the following form, $[\hat{\rho}_2(N)]_{ij} = c_{0; ij} + c_{1; ij} N^{-c_{2; ij}} e^{-\Delta_{ij} N}$, where $c_{0,1,2; ij}, \Delta_{ij}$ with $i,j=1,2,3,4$ are non-universal constants. 
The convergence of $\hat{\rho}_2(N)$ is slowest for the cases with conformal symmetries. For example, away from $g=1/2$, the TFIM shows faster convergences.  
In Fig.\,\ref{reduced matrix convergence}, we illustrate $\Delta_{11}$ of in TFIM. 
Since machine learning is very effective with any pattern recognition, we believe the pattern of $\hat{\rho}_2(N) $ is one of the reasons why our algorithm with machine learning is effective.

In conclusion, we have demonstrated that our machine learning-based algorithm can prepare the ground states of large systems. For example, the ground state of TFIM for a system size of 40 can be prepared with a small error ($\simeq 0.414\%$). Furthermore, our research has also shown that our algorithm can effectively generate the ground states of the XY and XXZ models for large systems. In a general sense, due to the broad applicability of our algorithm beyond the states considered, the results of our research support that our algorithm is a feasible and practical approach for preparing non-trivial ground states of interest for large systems.

Our algorithm can be used for quantum simulations in the near term, such as ion traps and Rydberg atoms. Hybrid algorithms have already been successfully applied to one-dimensional trapped ion problems \cite{kokail2019self}, and potential applications in scenarios involving a small number of qubits have been proposed \cite{ho2019efficient,zhou2020quantum}. Given that our algorithm is an extension of these hybrid algorithms, we believe it holds substantial applications and can offer significant benefits. Implementing our algorithm could be a crucial milestone in simulating highly entangled states with large numbers of qubits.

{\it Acknowledgements:} We thank Soonwon Choi and Taeckyung Lee for fruitful discussions. This work was supported by National Research Foundation of Korea under the grant numbers, NRF-2021R1A2C400847 and NRF-2022M3H4A1A04074153, as well as National Measurement Standard Services and Technical Services for SME funded by Korea Research Institute of Standards and Science (KRISS-2022-GP2022-0014).

%

	
\end{document}